\documentclass[aps,article,a4paper,10pt,notitlepage,footinbib,superscriptaddress,longbibliography,preprintnumbers]{revtex4-1}
\usepackage[english]{babel}
\usepackage{lmodern}
\usepackage[applemac]{inputenc}
\usepackage{amssymb,amsmath,amsfonts}
\usepackage{textcomp}
\usepackage{braket}
\usepackage{ulem}
\usepackage{soul}
\usepackage{graphicx}
\usepackage{dcolumn}
\usepackage{bm}
\usepackage{xcolor}
\usepackage{soul}
\usepackage{hyperref}
\usepackage{comment}
\usepackage{float}
\usepackage{soul}
\usepackage{orcidlink}

\def\I{\bm{I}_2}
\def\p{\textbf{p}}
\def\n{\textbf{n}}
\def\ap{\alpha_{\p}}
\def\kp{k_{\p}}
\def\kn{k_{\n}}
\def\mup{\bm{\mu}_{\bm{p}}}
\def\enne{\mathbf{n}}
\def\tnat{\tau_\text{nat}}
\def\lnat{\ell_\text{nat}}

\begin{document}

\title{Ordering and Defect Dynamics in Passive and Active Nematopolars}

\author{Fabio Aprile~\orcidlink{0009-0007-3388-7946}}
\affiliation{Dipartimento Interateneo di Fisica, Universit\'a degli Studi di Bari, via Amendola 173, Bari, I-70126, Italy}
\affiliation{Istituto Nazionale di Fisica Nucleare, Sezione di Bari, via Amendola 173, Bari, I-70126, Italy}

\author{Massimiliano Semeraro~\orcidlink{0000-0001-8273-4232}}
\affiliation{Laboratoire de Physique Th{\'e}orique et Mod{\'e}lisation, CNRS UMR 8089, CY Cergy Paris Universit{\'e}, F-95032 Cergy-Pontoise Cedex, France}

\author{Giuseppe Gonnella~\orcidlink{0000-0002-1829-4743}}
\affiliation{Dipartimento Interateneo di Fisica, Universit\'a degli Studi di Bari, via Amendola 173, Bari, I-70126, Italy}
\affiliation{Istituto Nazionale di Fisica Nucleare, Sezione di Bari, via Amendola 173, Bari, I-70126, Italy}
\affiliation{Kavli Institute for Theoretical Physics, University of California Santa Barbara, Santa Barbara, CA 93106, USA}
\preprint{BARI-TH/790-26}

\begin{abstract}
The coexistence of polar and nematic interactions, observed in a broad range of biological and synthetic active systems, gives rise to a rich phenomenology that continues to challenge our theoretical understanding of non-equilibrium collective behaviour.
In this paper, we numerically investigate phase ordering and defect dynamics in a newly introduced minimal single-field model for dry nematopolar systems, where competing polar and nematic contributions enter the free energy, and activity is implemented through self-advection. 
At optimal balance, the system develops depolarization strings connecting half-integer defects and separating domains with opposite polarization, together with closed depolarization loops.
We first characterize the elementary relaxation mechanisms of defect pairs and loops, showing that the interplay between polar and nematic alignment gives rise to non-monotonic string-mediated interactions, finite equilibrium separations and distinct loop-collapse pathways.
Large-scale simulations from disordered states instead show dynamic scaling with a characteristic length growing as $\sim(t/\ln t)^{1/2}$, consistent with coarsening in systems with non-conserved order parameters and point-like defects.
Upon introducing self-advection, sufficiently strong activity leads to the coexistence of positive integer and negative half-integer defects, which we term motility-induced charge symmetry breaking, and to saturation of the characteristic length scales, ultimately resulting in arrested coarsening. 
Overall, our results provide a simple unified framework for understanding the ordering and defect dynamics in biological and synthetic nematopolar systems.
\end{abstract}

\maketitle

\section{Introduction}

Coarsening dynamics defines a class of out-of-equilibrium processes in which ordered domains grow in size over time following universal scaling laws set by the symmetries of the system and by the conservation (or not) of the order parameter \cite{bray1991, bray1994}. 
Coarsening is mirrored by the growth of correlations with time and can be quantified by the behaviour of a characteristic length scale $L(t)$.
For non-conserved scalar order parameters, i.e. model A-like dynamics\cite{hohenberg1977}, coarsening is driven by curvature, leading to a domain size growth $L(t)\sim t^{1/2}$\cite{bray1994, cugliandolo2015}. 
When the system instead exhibits continuous rotational symmetry, as in $O(n)$ models \cite{pelissetto2002}, the dynamics is strongly influenced by the emergence and interaction of {\it topological defects}. 
For $n=2$, corresponding to the well-known XY model \cite{kosterlitz1974}, point-like defects lead to the growth law $L(t)\sim(t/\ln t)^{1/2}$ \cite{bray2000}. 
This scaling is reflected in the behaviour of the correlation functions, whose decay is influenced by the characteristic defect separation and becomes progressively slower as defects annihilate during coarsening.
Remarkably, this behaviour depends only on the dimensionality of the defect structure \cite{dutta2005Bis}, and therefore also applies to two-dimensional nematic systems, characterized by an $O(2)/\mathbb{Z}_2$ symmetry and featuring again point-like defects \cite{zapotocky1995, dennison2001, dutta2005Bis}. 

The aforementioned cases involve systems characterized by a single order-parameter symmetry, either nematic or polar, i.e. head-tail symmetric or not. 
A natural extension is to consider systems in which these two symmetries coexist. 
In such {\it nematopolar} systems, topological defects of both integer and half-integer charge may interact, leading to non-trivial spatial organization constrained by topology \cite{paik2026}. 
Blended nematopolar symmetry arises in a variety of physical systems. A paradigmatic example is provided by ferroelectric liquid crystals \cite{chen2020, lavrentovich2020, basnet2022, kumari2023, ma2024}, composed of rod-shaped molecules with a non-zero electric dipole moment, which, upon cooling, develop polar domains separated by material-dependent domain walls. 
Active realizations are even richer. These include, for example, living liquid crystals \cite{zhou2014, genkin2017, sokolov2015, Turiv2020}, where self-propelled polar particles embedded in a nematic background align with the director and form localized polarized regions.
Similar behaviour is observed in systems of polar constituents, such as bacteria \cite{volfson2008, doostmohammadi2016, meacock2021, wheeler2024, han2025} and eukaryotic cells \cite{saw2017, kawaguchi2017, blanch2018, ruider2024, ma2026}, which can display effective nematic order on large scales.
Microtubule–motor mixtures \cite{kruse2005, sumino2012, huber2018, roostalu2018} provide another key example, as tuning system parameters can lead to polar, nematic, or intermediate regimes where both types of order coexist. 

Capturing theoretically the relevant phenomenology of nematopolars requires suitable {\it minimal models}. 
In recent years, several approaches have been proposed to describe systems with mixed symmetry \cite{venkatesh2025}, ranging from particle-based models, such as self-propelled rods \cite{baskaran2008, peshkov2012, marchetti2013} and active dumbbells \cite{schwarz2012, cugliandolo2017, clopes2022, carenza2025}, to lattice theories \cite{lee1985, mondal2024} and continuum field theories.
In the continuum framework, one can distinguish between models with two different $n$-atic order parameters, where a polar field ($n=1$) is coupled to a nematic tensor ($n=2$) \cite{vats2024, vafa2025, mishra2025, dinelli2026}, and single-field models, in which a polar field, often coupled to a hydrodynamic velocity field, generates an effective nematic interaction \cite{amiri2022, ma2026}.

While the coarsening behaviour of two-field models has been extensively studied \cite{mishra2025}, existing work with single-field models has so far focused primarily on active turbulence rather than on phase ordering dynamics \cite{amiri2022}. 
Although systems in which polar and nematic behaviour originate from distinct components, such as living liquid crystals \cite{zhou2014, genkin2017, sokolov2015, Turiv2020}, are more naturally described by two-field models, single-field approaches are particularly well suited for intrinsically nematopolar systems, such as ferroelectric liquid crystals \cite{chen2020, lavrentovich2020, basnet2022, kumari2023, ma2024} and microtubule-motor complexes \cite{kruse2005, sumino2012, huber2018, roostalu2018}, where both types of order emerge from the same degrees of freedom.
However, coarsening properties of single-field models remain largely unexplored, and a systematic characterization of coarsening properties and growth laws is still lacking, especially in the presence of self-propelled contributions to the dynamics.

In this paper, we propose and numerically study a minimal single-field dry model for nematopolar systems, capable of reproducing their characteristic phenomenology while also providing a suitable framework to investigate dynamics, coarsening and ordering.
The model is based on a {\it single vector field} evolving through purely relaxational dynamics, with local interactions favoring simultaneous polar and nematic alignment via distinct contributions to the free energy. 
{\it Activity} is introduced through a self-advection term which, in the spirit of Toner–Tu theories for dry active matter \cite{toner1995}, allows the polarization to be interpreted simultaneously as an order parameter and as a local velocity field.
In the limits of vanishing nematic or polar alignment, the model consistently recovers vectorial descriptions of purely polar or nematic systems, respectively, although the latter are more commonly formulated in tensorial terms. 
In this respect, we remark that our model does not represent a generalization of the usual equilibrium liquid crystal theory \cite{degennes1993}, but rather a well-posed and useful approach specifically formulated to study dynamical properties of systems showing mixed nematopolar symmetries.

Despite its remarkable simplicity, our model is able to capture a rich defect phenomenology, thus enabling to clearly disentangle the individual roles of competing symmetries and activity in determining ordering kinetics.
Indeed, the numerical study of the ordering dynamics, for an optimal balance between polar and nematic alignment, shows a defect proliferation together with domain walls separating regions of opposite polarization.
Resorting to controlled simulations, we first analyze the evolution of minimal singular configurations emerging during the system evolution: half-integer defects connected by depolarization lines, or {\it strings} \cite{ma2024}, of vanishing polarization.
Defects move along these strings, which provide favourable pathways to reduce the overall energy of the system, with oppositely charged defects {\it annihilating} and like-charged ones reaching a {\it finite equilibrium separation}. 
We also observe the formation of closed strings, or {\it loops}, of vanishing polarization, which are observed to disappear according to two different mechanisms: 
either they progressively decrease in size, or the polarization field within them continuously rotates, in some cases leading to a loop rupture.
In all the above cases, we show that dynamical evolution is strongly influenced by competition between polar and nematic characters of the system. 
We then study the ordering process of large configurations starting from random initial conditions.
Correlation functions reveal that the growth law $L(t)\sim (t/\ln t)^{1/2}$ remains satisfied, and this result is confirmed by the analysis of the defect count. 
Finally, we turn on advection, which makes the system effectively active. 
We find that sufficiently strong self-advection leads to a {\it motility-induced charge symmetry breaking}, characterized by the coexistence of negative half-integer and positive integer defects (aster-like).
After a while, the system reaches a stationary configuration characterized by a near-constant number of defects and plateauing typical lengths, which can be rationalized in terms of dynamical stability of defect structures and mark the emergence of an {\it arrested coarsening regime}.

The remainder of the paper is organized as follows.
In section~\ref{sec:model_num_meth} we detail our model and numerical methods, and outline our main observables of interest. 
In section~\ref{sec:overview} we present an overview and a qualitative analysis of the singular structures that appear during the coarsening dynamics.
We then move on to the presentation and comment of our results. 
In particular, in section~\ref{sec:isolated_def} we deal with controlled configurations of two defects connected by a string and of loops, in section~\ref{sec:ord_dyn_full} we describe the nematopolar phenomenology of large systems and study how it is affected by different levels of relative polar and nematic alignment strengths, and in section~\ref{sec:advection} we show how the overall picture is modified by the inclusion of self-advection. 
Finally, in section~\ref{sec:conclusions} we briefly discuss the impact of our findings and draw the conclusions of our investigation.

\section{Model and Methods}
\label{sec:model_num_meth}

We consider a continuum model based on a polar vector field $\mathbf{p}$, which at the same time supports the emergence of topological defects with both integer and half-integer charges and captures the ability of self-propelled matter to move along a preferred direction.

The equilibrium properties of our system are ruled
by the following free energy functional
\begin{equation}
    \begin{split}
    F[\p] = 
    \int d^d x \,\bigg[&
    -\ap\left( -\frac{|\p|^4}{4} + \frac{|\p|^2}{2}\right)
    + \frac{\kp}{2}|\nabla \p|^2
    + \frac{\kn}{2}\big|\nabla\hat P\big|^2 \bigg]~,
\label{eq:free_energy}
\end{split}
\end{equation}
where position and time dependencies are implicit. The first term is a double-well potential which encodes the bulk behaviour of the polar field, as it sets the polarization equilibrium value at $p\equiv|\p|=1$. 
The second term $(\nabla\p)^2 \equiv \sum_{ij}(\partial p_i/ \partial r_j)^2$ ($i,j = 1,2$ dimensional indices) accounts for spatially inhomogeneous deformations and corresponds to the single elastic constant approximation.
This term penalizes any configuration rather than direction alignment of $\p$ vectors, with energy cost controlled by $\kp\geq0$. 
Finally, the last term represents a nematic-like contribution which favours orientation alignment of neighbouring vectors irrespective of their reciprocal direction, with energy cost controlled by $\kn\geq0$. 
We remark that here
\begin{equation}
    \hat P \equiv \left(\p\p^\top-|\p|^2\frac{\I}{2} \right)
\label{eq:nematic_tens}
\end{equation}
provides a tensorial representation of the nematic orientational order in terms of the polarization field $\p$. 
We remark that Eq.~(\ref{eq:nematic_tens}) gives the structure of the usual $\hat Q$ tensor of the liquid crystals theory \cite{degennes1993} upon performing the identification $S\equiv p^2$, $p\equiv |\p|$ being polarization magnitude, with $S$ restricted to be the positive scalar amplitude of $\hat Q$.

The dynamics of the system are governed by the following vector equation (see appendix~\ref{app:adim_eqs} for an adimensional formulation)
\begin{equation}
    \dot{\p} + \Lambda (\p \cdot \nabla)\p = -\Gamma\mup~,
    \label{eq:p_equation}
\end{equation}
where $\Lambda$ controls the strength of polarization self-advection, $\Gamma$ is the rotational viscosity and $\mup$ represents the molecular field, which, reflecting the structure of the free energy Eq.~(\ref{eq:free_energy}), can be decomposed into bulk, polar and nematic contributions as
\begin{equation}
    \boldsymbol{\mu}_{\p} \equiv \frac{\delta F}{\delta \p}
    = \boldsymbol{\mu}_{\p}^{\mathrm{bulk}}
    + \boldsymbol{\mu}_{\p}^{\mathrm{nem}}
    + \boldsymbol{\mu}_{\p}^{\mathrm{pol}}~,
\end{equation}
with
\begin{equation}
\begin{split}
    \mup^{\mathrm{bulk}} &= -\,\alpha_{\p}\big[1 - p^2 \big]\p~,\\
    \mup^{\mathrm{pol}}  &= -\,\kp\,\nabla^2 \p~,\\
    \mup^{\mathrm{nem}}  &= -\,2\kn\left[
                            p^2\,\nabla^2 \p
                            + 2\,(\nabla\p)^T(\nabla\p)\cdot \p
                            -\frac{2}{d}|\nabla \p|^2\p
                            -\left(1-\frac{2}{d}\right)(\p\cdot \nabla^2\p)\p  \right]~.
\end{split}
\end{equation}
The non-equilibrium character of the model emerges for $\Lambda\neq0$, when the advective term drives the system out of equilibrium and makes it effectively active.
In the following we focus on the two-dimensional case $d=2$, hence the term $(\p\cdot \nabla^2\p)\p$ in $\mup^{\mathrm{nem}}$ will not be present in our equations. 
Additionally, due to the properties of symmetric and traceless $2\times2$ tensors, as $\hat Q$ and $\hat P$, no contribution corresponding to the usual liquid crystal term $\text{Tr}(\hat Q^3)$ is required in the free energy Eq.~(\ref{eq:free_energy}) in this dimensionality.
We remark that $\Lambda\downarrow 0, k_{\p}\downarrow 0$ represents the passive and nematic-symmetric limit of our model, in which it loses its active self-advection contribution and becomes invariant under the inversion $\p\rightarrow-\p$.

\subsection{Numerical integration and observables of interest}
\label{sec:num_meth}

We integrate equation~(\ref{eq:p_equation}) resorting to a finite-difference scheme, where differential operators are computed by standard stencil techniques \cite{leveque2007}.
Simulations are run for $\sim 10^6$ iterations applying periodic boundary conditions on two-dimensional square lattices of linear size $N$ taking values $512$ and $1024$, unless otherwise specified.
Unless stated otherwise, we fix lattice spacing to $\Delta N=1$ and time step to $\Delta t=10^{-2}$, verifying that our results are numerically stable when decreasing $\Delta N$ and $\Delta t$. 
We present our results in simulation units: time and space are discretized as $t= n_i\Delta t$, with $n_i$ the iteration index, and $x= n_x\Delta N,~
y= n_y\Delta N$, with $n_x,n_y$ horizontal and vertical node indices, giving $L=N\Delta N$ for system dimension.
We will make occasional use of time and length natural units, $\tnat\equiv1/(\Gamma \alpha_{\p})$ and $\lnat\equiv\sqrt{(\kp+2\kn)/\alpha_\mathbf{p}}$, to rescale our data. 
These respectively emerge from the adimensional form of equation~(\ref{eq:p_equation}) (see appendix~\ref{app:adim_eqs}) and an approximate solution for polarization profile sufficiently far from defect cores (see appendix~\ref{app:pol_dec}).

For large scale simulations, we initialize the system in a disordered state, i.e. with $\p$ having unitary modulus and random orientation at each lattice node.
Controlled simulations of pairs of half-integer defects and loops are instead started from ad-hoc initial conditions, as detailed in section~\ref{sec:isolated_def}. 
The parameter values are fixed to $\Gamma=1$, $\alpha_{\p}=0.1$ (giving $\tnat=10$) and $\Lambda=0.1$ (when advection is active). 
For comparison purposes, we also define the adimensional number $\bar k \equiv \kp/\kn \geq 0$ (see appendix~\ref{app:adim_eqs}).
We explore different relative polar and nematic strengths varying $\bar k$ acting on $k_{\n}$ while keeping $k_{\p}$ fixed (although we checked that similar qualitative results are obtained varying $k_{\p}$ while keeping $k_{\n}$ fixed).

Observables of interest for our investigation are the location, number and type of topological defects, and the polar and nematic correlation functions of $\p$. 
Following \cite{hobdell1997}, the former are computed by employing the {\it winding angle approach}. 
In practice, for each lattice node we consider a plaquette for which the node of interest is at the bottom-left corner and evaluate the rotation of $\p$ (modulo $\pi$ for half-integer defects) along its edges.
Defects are then identified from the corresponding non-null winding numbers, making sure that the vanishing-charge topological constraint is always satisfied.
The normalized polar \cite{liu1990, bray1991} and nematic \cite{zapotocky1995, bhattacharjee2008} two-point correlation functions
\begin{equation}
    C_{\p}(r,t) = \frac{ \braket{\p(0,t)\p(r,t)}}{\braket{\p(0,t)\p(0,t)}}\qquad\text{and}\qquad
    C_{\n}(r,t) = \frac{ \braket{\hat P(0,t)\hat P(r,t)}}{\braket{\hat P(0,t)\hat P(0,t)}}
\label{eq:corr_funs}
\end{equation}
are computed using a fast Fourier transform approach, followed by radial averaging to obtain isotropic functions of the distance $r$.
Moreover, as typically done \cite{mondello1990, dutta2005}, we also extract typical polar $L_{\p}(t)$ and nematic $L_{\n}(t)$ correlation lengths by observing how the distance at which the correlation functions reach a specific value changes over time. 
In other words, we examine $C_{\p}(L_{\p}(t),t)=c_{\text{p}}$ and $C_{\n}(L_{\n}(t),t)=c_{\text{n}}$ at fixed $c_{\text{p}}=c_{\text{n}}=0.2$.

\section{General features of nematopolar ordering}
\label{sec:overview}

In this section, we introduce and briefly discuss the most relevant  features of the defect phenomenology emerging during the ordering dynamics of our nematopolar system. 
Interestingly, the nematopolar landscape departs from the conventional picture according to which, in systems with unique well-defined symmetries, topological defects are point-like \cite{mermin1979, bray1994}, and are assigned a topological charge determined by the winding of the order parameter around their cores (integer or half-integer for $O(2)$ or $O(2)/\mathbb{Z}_2$, corresponding to polar or nematic symmetry, respectively). 
In contrast, during the ordering dynamics of the nematopolar system, relaxation toward an ordered state is not governed solely by the motion and annihilation of point-like defects, but rather the system develops a richer morphological landscape in which defects coexist with extended singular structures spanning finite regions of space. 
These peculiar structures emerge spontaneously during coarsening, reflect the competition between polar and nematic contributions to the ordering process and are consistent with similar structures reported in previous nematopolar studies \cite{amiri2022, vats2024, vafa2025, mishra2025, dinelli2026, ma2026}, which however often include multiple fields and explicit complex couplings.

\begin{figure}[t]
    \centering
    \includegraphics[width=0.9\linewidth]{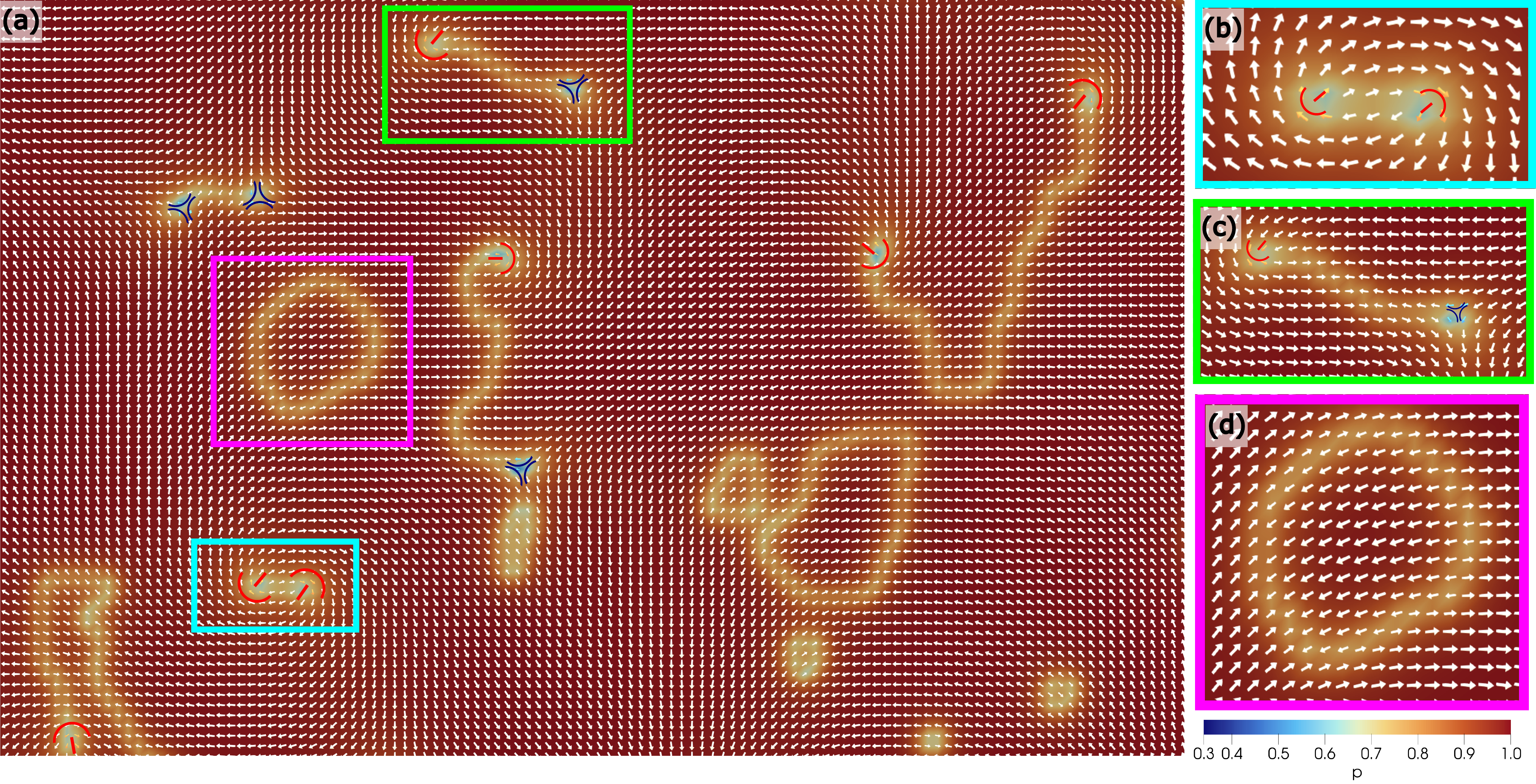}
    \caption{
    {\it A typical nematopolar configuration with string and loop structures.}
    $(a)$ Magnified view of a representative nematopolar configuration at $t=10^3$ for the case $\bar{k}=0.3$, started from a disordered initial state. 
    $(b)$-$(d)$ Defect structures emerging during evolution: a string connecting two like-charged half-integer defects $(b)$, a string connecting oppositely charged half-integer defects $(c)$ and a closed loop $(d)$. 
    These panels show enlargements of the regions delimited by matching rectangles in $(a)$. 
    In all panels, arrows indicate the local orientation of $\mathbf p$, while the background reports its local magnitude $p$ and is coloured according to the scale bar below. 
    Additionally, red comets and blue trefoils denote the location of $+1/2$ and $-1/2$ defects, respectively.
    Other parameters are given in section~\ref{sec:num_meth}.
}
\label{fig:fig1}
\end{figure}

As a visual representation of these features, we present in figure~\ref{fig:fig1}(a) a representative configuration for $\bar{k}=0.3$ ($\kp=0.03$, $\kn=0.1$). 
This value of $\bar{k}$ provides a particularly clear illustration of the characteristic nematopolar phenomenology. 
Indeed, at intermediate times, it displays the spontaneous emergence of extended singular
structures, while at late times the balance between $\kp$ are $\kn$ is such that integer and half-integer charge defects can simultaneously coexist (see figure~\ref{fig:fig5}).
For these reasons, we will focus on configurations with $\bar{k}=0.3$ for comparisons throughout the following discussions.

More in detail, the figure displays the emergence of strings in which pairs of half-integer defects -- whose location is marked by red comets ($+1/2$) and blue trefoils ($-1/2$) -- are connected by strings \cite{lee1985} -- yellow curvilinear structures.
Defects are characterized by the same (figure~\ref{fig:fig1}(b)) or opposite charges (figure~\ref{fig:fig1}(c)). 
The strings correspond instead to depolarization lines that separate regions with opposite polarization. 
These structures can be viewed as elongated integer-charge objects, with an overall topological charge that is unitary or vanishing, depending on the sign of the half-integer defects connected by the string \cite{vafa2025, dinelli2026}.
We also observe the formation of closed loop structures. 
These emerge in regions where an enclosed area of the system is characterized by a polarization opposite to that of its immediate surroundings (figure~\ref{fig:fig1}(d)).
Also here, at the interface between oppositely-oriented regions a closed depolarization line emerges.
In this respect, loops can be interpreted as strings connecting oppositely charged defects that have closed onto themselves, thereby annihilating into a charge-neutral structure \cite{vafa2025, dinelli2026}.

The tendency of our nematopolar system to generate these string and loop structures can be rationalized as follows. 
During the early stages of dynamics, the system rapidly evolves from a disordered initial condition into configurations populated by numerous half-integer defects, which are energetically preferred over their integer counterparts due to their lower elastic energy.
By definition, the order parameter rotates by $\pi$ around their cores, implying that, in the vicinity of each half-integer defect, there exist locations where the polarization vectors point in opposite directions. 
For example, in figure~\ref{fig:fig1}(b) and (c) these regions are located at the tail of the comet-like structure for the positive defect and at one of the trefoil arms for the negative one.
Since here the polarization field interpolates between opposite directions, its magnitude across these walls is locally reduced, giving rise to depolarization strings connecting defects.
Once formed, defects move along such strings to lower the overall energy. 
Depending on the charges of the connected defects, these either annihilate or, in absence of interaction with other defect structures, relax to a stationary equilibrium separation, as discussed in detail in section~\ref{sec:defects_strings}.
Closed depolarization loops instead arise either during the initial ordering process or subsequently as a consequence of string dynamics. 
In this case, system lowers its energy through loop elimination. As discussed in section~\ref{sec:loops}, we identify two different mechanisms by which the orientational discontinuity is resolved: 
i) evaporation due to a progressive size decrease;
ii) reorientation of the polarization field inside the loop, which can eventually lead the loop to rupture into different charge-integer strings.

\section{Dynamics of isolated defect and loop structures}
\label{sec:isolated_def}

Having commented on the general phenomenology of our nematopolar model, we now discuss in more detail the properties and evolution of the emerging defect and depolarization structures. 
To do so, we resort to controlled simulations started from ad-hoc initial conditions built including either a pair of half-integer defects connected by a string or a circular depolarization loop.

\subsection{Pairs of half-integer defects connected by a string}
\label{sec:defects_strings}

We first investigate pairs of half-integer defects connected by straight domain walls separarting regions of opposite polarization. 
The system is initialized with two defects, either like or oppositely-charged, positioned at an initial separation length $\ell_0$ along the horizontal axis of the system. 
The polarization profile across the domain wall was observed to relax very soon to a smooth interface without appreciable changes in the position of the defect cores, and with a depolarization line along the interface. 
We consider systems with linear size $N=256$, set $\ell_0\sim N/3$, and checked that our findings remain qualitatively unchanged for different initial separations. 

\begin{figure}[t!]
    \centering
    \includegraphics[width=0.9\linewidth]{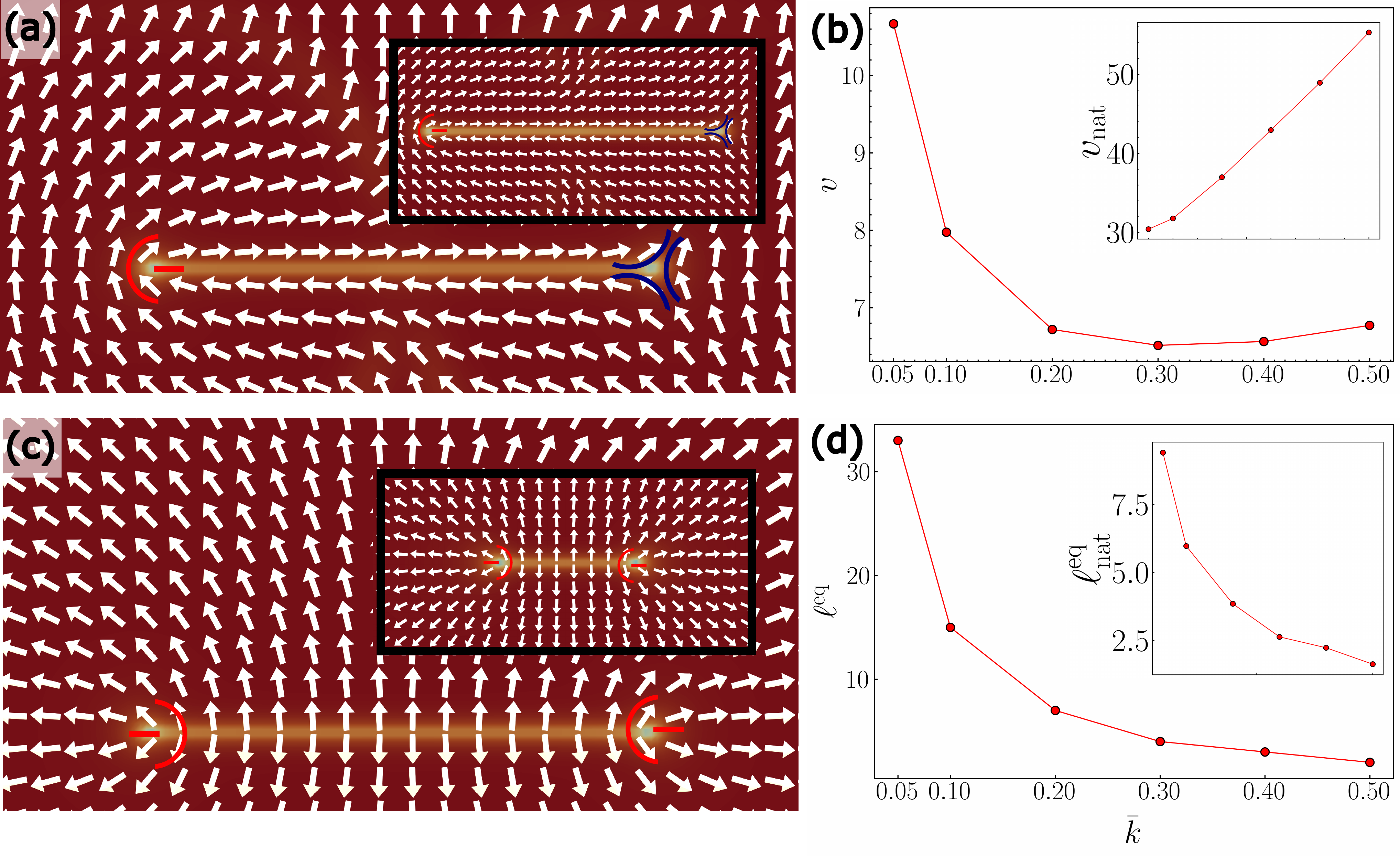}
    \caption{
    {\it Pairs of half-integer defects in controlled settings.}
    $(a)$ Magnified view of a pair of oppositely charged defects evolved in the case $\bar{k}=0.3$ and connected by a string. 
    The corresponding far-field configuration displays an ordered arrangement, with arrows all pointing upwards.
    Inset: same defect pair evolved in the case $\bar{k}=0.5$.
    $(b)$ Annihilation velocity $v$ of oppositely charged defect pairs as a function of $\bar{k}$. $v$ is estimated as the ratio between the initial separation $\ell_0$ (identical in all cases) and the time required for the defects to travel along the string and come into contact.
    Inset: dimensionless annihilation velocity $v_{\rm nat}\equiv v(\tnat/\lnat)$, expressed in natural units.
    $(c)$ Pair of positive like-charged defects.
    As in $(a)$, defects are evolved in the case $\bar{k}=0.3$ and are connected by a string. 
    Differently from $(a)$, here the far-field configuration displays an aster-like profile.
    Starting from their initial positions (main panel), the defects move along the string until they reach a stationary equilibrium separation (inset).
    $(d)$ Equilibrium distance $\ell^{\rm eq}$ between two positive like-charged defects as a function of $\bar{k}$. 
    Inset: dimensionless equilibrium distance $\ell^{\rm eq}_{\rm nat}\equiv \ell^{\rm eq}/\lnat$, expressed in natural units.
    $(a)$ and $(c)$ follow the same representation style as in figure~\ref{fig:fig1}.
    In all cases, simulations are performed on a lattice linear of linear size $N=256$, initial conditions are prepared as described in section~\ref{sec:defects_strings}, and defects are initially placed at an initial separation $\ell_0\sim N/3$.
    Other parameters are given in section~\ref{sec:num_meth}.
}
\label{fig:fig2}
\end{figure}

In figure~\ref{fig:fig2}(a) we report a typical configuration at initial times for a couple of $+1/2$ and $-1/2$ defects in the case $\bar{k}=0.3$. 
Although oppositely-charged defects are known to feel a Coulomb-like attraction \cite{kosterlitz2016, harth2020}, as we will show in a moment, the connecting string provides a privileged pathway to lower system energy along which defects approach. 

In figure~\ref{fig:fig2}(b), we report the average annihilation velocity $v$ as a function of $\bar k$. 
This is estimated as the ratio between the initial separation $\ell_0$ and the time required for the defects to come closer along the string and annihilate.
Interestingly, $v$ features a non-monotonic trend, which can be rationalized in terms of two competing effects. 
On the one hand, a straightforward extension of 
a classical argument
\cite{kosterlitz1973} gives that the interaction energy between two opposite charge defects is
\begin{equation}
    E_{\mathrm{int}}\sim-\left(k_{\p}+2k_{\n}\right)\ln\frac{\ell}{\ell^{\mathrm{c}}}~,
\label{eq:def_int_en}
\end{equation}
with $\ell$ actual defect separation and $\ell^{\mathrm{c}}$ core size.
Therefore, decreasing $k_{\n}$ while keeping $k_{\p}$ fixed ($\bar k$ increases), reduces the strength of the effective interaction and explains why for small $\bar k$ the velocity $v$ reduces. 
On the other hand, one has to take into account also the role played by the string.
As $k_{\n}$ decreases further, the polar character of the system is expected to become more relevant (see the inset of figure~\ref{fig:fig2}(a), showing a more marked sample string at $\bar k=0.5$) and a string energy, 
\begin{equation}
    E_{\mathrm{string}}\sim\sigma \ell g(\bar k)~,
\label{eq:def_int_string}
\end{equation}
has to be taken into account.
Here we approximate this quantity by factorizing the contribute of a domain wall in a pure polar system, given by $\sigma\sim\sqrt{\alpha_{\p}k_{\p}}$, and introducing the function $g(\bar k)$ that incorporates the presence of the nematic field and that can be interpreted as an effective screening. 
The function $g(\bar k)$ would vanish at $\bar k=0$ (for which no string emerges) and would increase monotonically until a value of order $1$. 
Therefore, at larger $\bar k$, the string energy contribution becomes more effective, and defects can be expected to accelerate their motion along the string to reach a lower energy configuration more rapidly. 
This argument could explain the overall non-monotonic trend observed for $v$.
We note that contributions coming from the far field regions, scaling as $\ln(\ell^{\mathrm{sys}}/\ell^{\mathrm{c}})$, with $\ell^{\mathrm{sys}}$ system extension, bear no $\ell$ dependence, and can be neglected in our arguments.

In order to disentangle the simultaneous action of the two dominant effects described above, we rescale $v$ in natural units as $v_{\mathrm{nat}} \equiv v(\tnat/\lnat)$, thus making it a dimensionless observable. 
This rescaling effectively factors out the dependence on interaction strength, so that the trend of $v_{\mathrm{nat}}$ is only affected by string-related effects.
The resulting curve, shown in the inset of figure~\ref{fig:fig2}(b), indeed exhibits a growing monotonic trend, consistent with the above interpretation of the role played strings: as $\bar k$ is increased, thus moving into configurations with a more marked polar character, defects tend to move faster along strings to lower system energy more rapidly. 

We now turn to the case of like-charged defects. 
In purely nematic systems, these are known to experience a Coulomb-like repulsive interaction \cite{kosterlitz2016, harth2020}. 
However, in the nematopolar setting, depolarization strings emerge, providing energetically favorable pathways that qualitatively modify the interaction. 
As a result, two competing tendencies, both driven by energy minimization, are again at play. 
On the one hand, the system tends to increase the separation between like-charged defects; on the other hand, it tends to reduce the length of the connecting string. 
As a consequence, defects move along the string until a balance between these two effects is reached at a stationary equilibrium separation.
This behaviour is illustrated in figure~\ref{fig:fig2}(c) for a pair of $+1/2$ defects in the case $\bar{k}=0.3$: starting from sufficiently separated initial positions (main panel), the defects approach each other along the string and eventually settle at an equilibrium distance $\ell^{\mathrm{eq}}$ (inset).
Vice versa, starting from sufficiently close locations, defects first repel and then settle at $\ell^{\mathrm{eq}}$.

From the perspective of string energetics, one may expect that increasing $\bar{k}$ raises the energy cost $E_{\text{string}}$ of the string and enhances the tendency of the system to shorten the string, thus decreasing $\ell^{\mathrm{eq}}$. 
This trend is confirmed in figure~\ref{fig:fig2}(d), where we report $\ell^{\mathrm{eq}}$ as a function of $\bar{k}$. 
A quantitative interpretation can be obtained by minimizing the total energy with respect to defect separation. 
As discussed above, relevant energy contributions are the interaction and string energies, given in equation~(\ref{eq:def_int_en}) and equation~(\ref{eq:def_int_string}), respectively.
Straightforward steps show that $E_{\mathrm{int}} + E_{\mathrm{string}}$ is minimized at
\begin{equation}
    \ell^{\mathrm{eq}} \sim \frac{k_{\mathbf p}+2k_{\mathbf n}}{\sigma g(\bar{k})} = \frac{\ell_{\mathrm{nat}}}{h(\bar{k})}~,
\label{eq:def_sep}
\end{equation}
where we used $\sigma \sim\sqrt{\alpha_{\p}k_{\p}}\ \sim \alpha_{\mathbf p}\ell_{\mathrm{nat}} f(\bar{k}) $, with $ f(\bar k)\equiv\sqrt{\bar k/(\bar k + 2)}$, and we defined $h(\bar{k}) \equiv f(\bar{k})g(\bar{k})$, which is an increasing function of $\bar{k}$. 
Interestingly, if $g(\bar{k})$ increases faster than $\sim \sqrt{\bar{k}}$, equation~(\ref{eq:def_sep}) becomes a decreasing (increasing) function of $k_{\mathbf p}$ ($k_{\mathbf n}$), in agreement with figure~\ref{fig:fig2}(d). 
Moreover, equation~(\ref{eq:def_sep}) naturally implies that, when expressed in natural units, thus again factoring out interaction strength, the dimensionless separation $\ell^{\mathrm{eq}}_{\mathrm{nat}} \equiv \ell^{\mathrm{eq}}/\lnat = 1/h(\bar{k})$ decreases with $\bar{k}$, as shown in the inset of figure~\ref{fig:fig2}(d).

\subsection{Depolarization loops}
\label{sec:loops}

We now turn our attention to depolarization loops. 
In this case, the initial condition is built by selecting a circular region of radius $R_0$ and assigning polar vectors of opposite directions in its interior and exterior. 
To better resolve the crossover between different loop relaxation mechanisms, simulations are performed using a reduced lattice spacing of $\Delta N=0.5$.
 The initial loop radius is fixed to $R_0\sim 26$ in a system of linear size $L=128$, while the system behaviour is explored by varying $\bar{k}$ and $\lnat$. 
The former controls the energetic cost of depolarization lines.
The latter instead regulates loop thickness, i.e. the typical length scale over which polarization magnitude decays from bulk value to zero across the loop from both its interior and exterior.

\begin{figure}[t]
    \centering
    \includegraphics[width=0.9\linewidth]{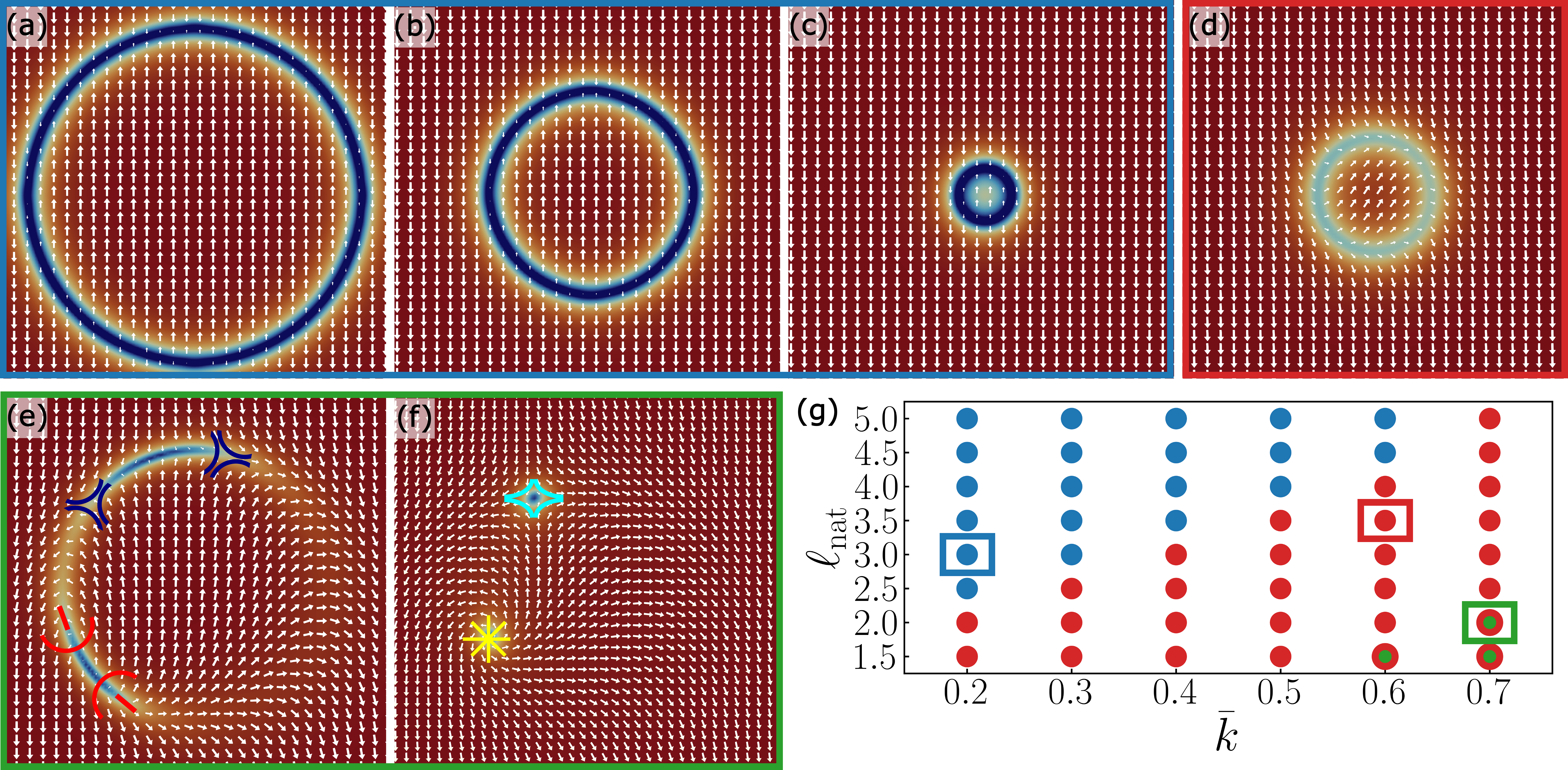}
    \caption{{\it Loop phenomenology: evaporation, rotational relaxation and rupture.}
    $(a)$ to $(c)$ Evolution of a depolarization loop at $(\bar k=0.2,\lnat=3.0)$. 
    Panels report successive snapshots of the gradual evaporation of the loop.
    $(d)$ Relaxation of a depolarization loop through a continuous rotation of the polarization field at $(\bar k=0.6,\lnat=3.5)$.
    $(e)$ to $(f)$ Rotational relaxation dynamics, followed by loop rupturing into two strings at $(\bar k = 0.7,\lnat=2.0)$.
    Strings initially connect a couple of positive and negative half integer defects and then contract, leaving a defect pair with charges $+1$ (yellow star) and $-1$ (cyan quatrefoil).
    $(g)$ Typical phase diagram of loop phenomenology in the $\bar k ~-~ \lnat$ plane. 
    Blue and red dots respectively denote the evaporation and rotational regimes, while green dots mark the emergence of rupture phenomena in the rotational regime.
    Matching rectangles mark the $(\bar k, \lnat)$ points at which snapshots in $(a)$ to $(f)$ were taken.
    In all cases, simulations are performed on a lattice of linear size $N=256$ and reduced lattice spacing $\Delta N=0.5$ (system dimension $L=128$), initial conditions are prepared as described in section~\ref{sec:loops}, and loops are initialized at the center of the system with radius $R_0=0.2~L\sim 26$.
    $(a)$ to $(f)$ follow the same representation style as in figure~\ref{fig:fig1}.
    Other parameters are given in section~\ref{sec:num_meth}.
}
\label{fig:fig3}
\end{figure}

We identify two distinct mechanisms through which the system resolves the orientational discontinuity and lowers its free energy.
The first consists of a progressive reduction of the loop size until its complete evaporation. 
According to equation~(\ref{eq:def_int_string}), the energy associated with the loop scales as
\begin{equation}
    E_{\mathrm{loop}} \sim
    \sigma \ell_{\mathrm{loop}} g(\bar{k}) \sim
    \sigma R g(\bar{k})~,
\label{eq:loop_energy}
\end{equation}
where $\ell_{\mathrm{loop}}$ is the loop circumference and $R$ its actual radius. 
Therefore, the loop energy decreases as the loop contracts. 
A representative realization of this process, which we call {\it evaporation mechanism}, is shown in figure~\ref{fig:fig3}(a)-(c) for $\bar{k}=0.2$ and $\ell_{\mathrm{nat}}=3.0$.
During the process, polarization vectors preserve their orientation inside and outside the loop, remaining parallel to the vertical axis while pointing in opposite directions.
The loop thus behaves as a closed domain wall for the polarization magnitude, and the mechanism of evaporation closely resembles the curvature-driven dynamics of interfaces in systems with a non-conserved scalar order parameter.

The second mechanism instead consists in removing the orientational discontinuity through a continuous rotation of the polarization field: in the close vicinity of the loop, polarization vectors tend to rearrange in such a way that the field orientation interpolates smoothly from the inside to the outside of the loop. 
A representative realization of this process, which we call {\it rotational mechanism}, is shown in figure~\ref{fig:fig3}(d) for $\bar{k}=0.6$ and $\ell_{\mathrm{nat}}=3.5$. 
Once reorientation has started, the system naturally evolves to an ordered configuration where the loop has completely disappeared.

Interestingly, upon increasing $\bar{k}$ and reducing $\lnat$, the loop can rupture into separate charged strings, which subsequently rearrange into integer-charge defects. 
A representative realization of this process is shown in figure~\ref{fig:fig3}(e)-(f) for $\bar{k}=0.7$ and $\lnat=2.0$.
Once the loop breaks, it gives rise to two strings, each carrying like-charged half-integer defects, consistent with the overall charge-neutral character of the original loop. 
Defects then approach along the strings, until eventually becoming indistinguishable from isolated integer defects with positive (yellow star) and negative (cyan quatrefoil) charge.

In figure~\ref{fig:fig3}(g) we report a phase diagram in the $(\bar k, \lnat)$ plane, where blue (red) dots mark the emergence of the evaporation (rotational) mechanism, while green symbols denote loop rupture.
The phase diagram structure can be understood in terms of the different mechanisms adopted by the system to resolve the orientational discontinuity associated with the loop.
Indeed, at fixed $\ell_{\mathrm{nat}}$, increasing $\bar{k}$ results in a change of mechanism from evaporation to rotational relaxation, which reflects a crossover from a nematically to a polar dominated alignment tendency.
More specifically, in the latter case, at larger $\bar{k}$, the system tends to remove the loop by continuously rotating the polarization field, so as to smoothly connect the orientations inside and outside the loop. 
Conversely, when $\bar{k}$ is decreased, the nematic interaction becomes dominant: antiparallel orientations in the interior and exterior of the loop are remain unaltered, so that a rotation of the polarization field becomes energetically less convenient. 
The system therefore resolves the discontinuity by preserving the direction of the field inside and outside the loop and progressively eliminating the depolarization line. 
This corresponds to the evaporation mechanism. 
Similar arguments also account for the lower-right region of the phase diagram. 
There, increasing $\bar{k}$ makes the reorientation across thin loops abrupt, eventually favouring loop rupture into charged strings.

Finally, to explain the behaviour observed at fixed $\bar{k}$ upon varying $\ell_{\mathrm{nat}}$, i.e. the increasing trend of the boundary separating the evaporation and rotational regimes, we provide an energetic argument comparing the cost of maintaining a loop with that of removing it through a reorientation of the polarization field.
The energy associated with a loop of radius $R$ is given in equation~(\ref{eq:loop_energy}). 
The energetic cost of the rotational pathway can instead be estimated as follows. 
During this process, the magnitude of the field remains approximately constant and close to its equilibrium value $p\simeq 1$, so that the dominant contribution arises from orientational degrees of freedom.
To estimate this contribution, we approximate the angular field as
\begin{equation}
    \theta(r) \sim \theta_0+\frac{\pi}{2}\left(1+\tanh\left(\frac{r/R-1}{\xi}\right)\right)~,
\label{eq:rotational_angle}
\end{equation}
with $r$ radial coordinate, $\theta_0$ orientation of the polar field at the centre of the loop and $R$ and $\xi$ loop radius and extension, the latter spreading for just a few simulations units. 
In agreement with our initial condition, equation~(\ref{eq:rotational_angle}) ensures that, from $r\sim 0$ to $r\gg R$, the polarization field rotates by $\pi$, with reorientation concentrated in the vicinity of the loop. 
Using equation~(\ref{eq:rotational_angle}), equation~(\ref{eq:grad_p_vec}), equation~(\ref{eq:grad_p_tens}), and equation~(\ref{eq:free_en_polar_coords}), the rotational energy can then be estimated as
\begin{equation}
    E_{\mathrm{rot}}\sim
    \frac{k_{\mathbf p}+2k_{\mathbf n}}{2}
    \int d^2x(\nabla \theta)^2 \sim
    k_{\mathbf p}+2k_{\mathbf n}~.
\end{equation}
The rotational pathway is therefore energetically favoured when $E_{\mathrm{rot}} < E_{\mathrm{loop}}$, which yields the condition
\begin{equation}
R > R_{\mathrm{c}} \sim \frac{k_{\mathbf p}+2k_{\mathbf n}}{\sigma g(\bar{k})}
= \frac{\ell_{\mathrm{nat}}}{h(\bar{k})}~,
\label{eq:critical_radius}
\end{equation}
where $R_{\mathrm{c}}$ denotes the critical radius.
This result reflects once again the competition between polar and nematic contributions.
When the nematic coupling dominates (small $\bar{k}$), the system favours configurations in which neighbouring vectors remain parallel or antiparallel, leading to larger $R_c$.
Conversely, for larger $\bar{k}$, the increased cost of depolarization lines makes the rotational pathway more favourable, resulting in a smaller critical radius.
Additionally, from equation~(\ref{eq:critical_radius}), one obtains that $\ell_{\mathrm{nat}} \lesssim R h(\bar{k})$. 
Since $h(\bar{k})$ is an increasing function of $\bar{k}$, and our simulations are performed at fixed $R$, this relation is consistent with the observed increase of $\ell_{\mathrm{nat}}$ as a function of $\bar{k}$ along the regime boundary in figure~\ref{fig:fig3}(g). 
Finally, equation~(\ref{eq:critical_radius}) also implies that the overall picture described above remains qualitatively valid for different initial radii, confirming the robustness of the phase diagram structure.

\section{Ordering dynamics, correlations and defect dynamics}
\label{sec:ord_dyn_full}

We are now in the position to analyze the evolution of the whole system, having in mind that typical configurations during the ordering process are similar to that shown in figure~\ref{fig:fig1}.

As discussed in section~\ref{sec:overview}, during the early stages of the dynamics the system rapidly evolves from a disordered initial condition into configurations containing numerous defects and orientationally-singular structures. 
In the polar case (diverging $\bar{k}$) these consist exclusively of integer-charge defects, whereas in nematopolar settings (finite $\bar{k}>0$) they consist of half-integer defects connected by strings, together with closed loops.
Subsequently, the system enters an ordering regime, i.e. a coarsening stage characterized by the progressive disappearance of defects and singularities.
In the polar case, ordering proceeds solely through the motion of integer defects driven by attractive or repulsive interactions ruled by relative charges, with oppositely charged pairs eventually annihilating upon contact.
In the nematopolar setting, closed loops either disappear, as described in section~\ref{sec:loops}, or are disrupted through interactions with nearby defects.
At the same time, half-integer defects move along connecting strings according to the mechanisms discussed in section~\ref{sec:defects_strings}: oppositely charged defects approach and annihilate, whereas like-charged ones reach an equilibrium separation. 
However, interaction effects with other nearby defects induce strings to collapse, so, if sufficiently close, like-charged half-integer defects effectively behave as isolated integer-charge ones, which subsequently are attracted to and annihilate with other integer-like defects of opposite charge in the system. 
If instead like-charged half-integer defects remain sufficiently far, connecting string are still disrupted, but now leave those far enough to be distinguishable. 
In this case, ordering dynamics locally proceeds as in usual nematic systems \cite{dutta2005, zapotocky1995}.
In any case, all defects and singular structures are eventually eliminated, and the system ultimately relaxes to a uniformly ordered stationary state.

\begin{figure}[t]
    \centering
    \includegraphics[width=0.9\columnwidth]{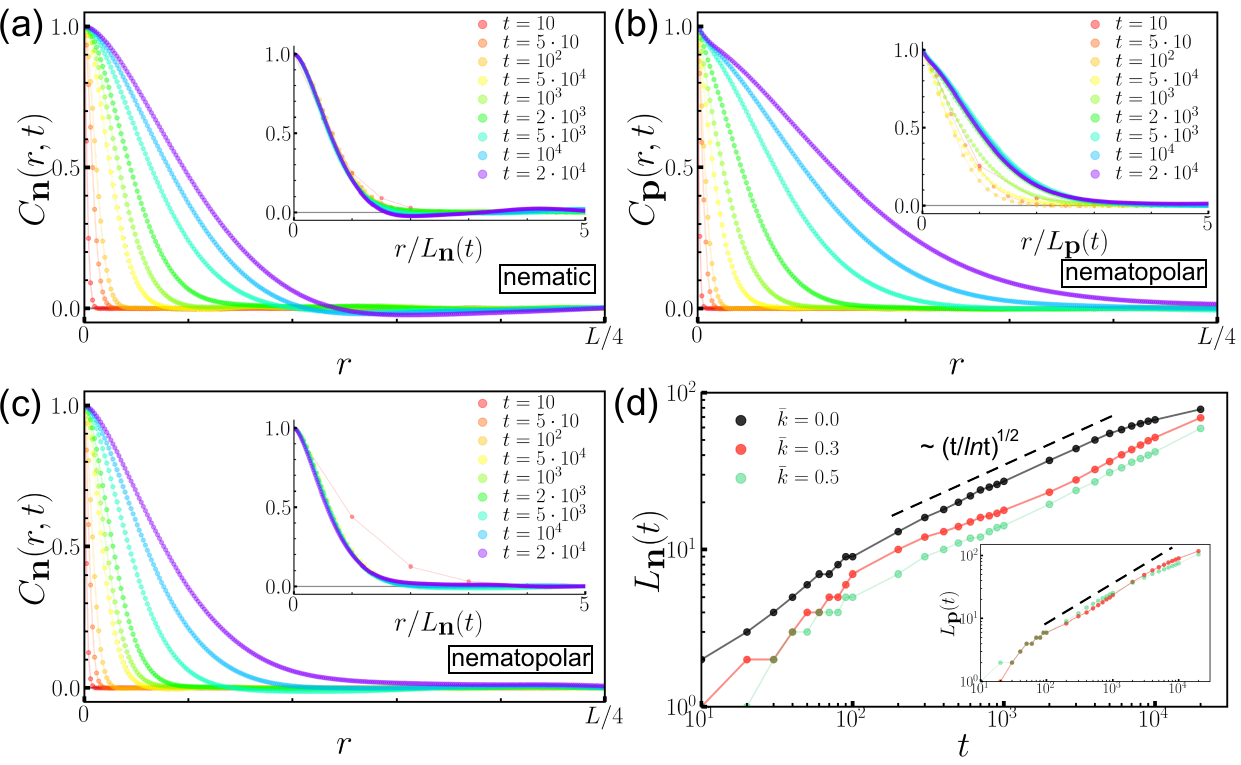}
    \caption{{\it Correlation functions and typical lengths.} 
    $(a)$ Nematic correlation function $C_{\n}(r,t)$ for the pure nematic case.
    $(b)$ and $(c)$ Polar and nematic correlation functions $C_{\p}(r,t)$ and $C_{\n}(r,t)$ for the case $\bar k=0.3$. 
    Insets report the curve collapse obtained by rescaling distances according to the relevant length scale $L_{\n}(t)$ or $L_{\p}(t)$ from $(d)$.
    $(d)$ Typical nematic $L_{\n}(t)$ (main) and polar $L_{\p}(t)$ (inset) lengths for different values of $\bar k$. 
    Dashed lines highlight the trend $\sim (t/\ln t)^{1/2}$. 
    All curves are averaged over $5$ independent runs. 
    Other parameters are given in section~\ref{sec:num_meth}.
}
\label{fig:fig4}
\end{figure}

This ordering dynamics is reflected in the behaviour of the polar and nematic correlation functions, defined in equation~(\ref{eq:corr_funs}).
As a reference, in figure~\ref{fig:fig4}(a) we report the nematic correlation function $C_{\n}(r,t)$ for the nematic case $\bar{k}=0$ at different successive times.
The corresponding polar correlation function is not shown, as it is trivially equal to unity at $r=0$ and vanishes for any $r>0$. 
As expected, the decay of $C_{\n}(r,t)$ becomes progressively slower with increasing time, reflecting the growth of orientational order as defects are eliminated.
In figure~\ref{fig:fig4}(b) and (c), we instead show the polar and nematic correlation functions, $C_{\p}(r,t)$ and $C_{\n}(r,t)$ measured in the case $\bar k = 0.3$ at the same times considered in panel (a). 
As expected, introducing a finite polar coupling $k_{\p}>0$ in the free energy equation~(\ref{eq:free_energy}) gives rise to a non-vanishing polar correlation function. 
The decay of $C_{\p}(r,t)$ thus becomes progressively slower at later times, consistently with the increasing degree of polar order attained by the system.
A comparison between the nematic correlation functions shown in figure~\ref{fig:fig4}(a) and (c) reveals that, although their qualitative behaviour is similar, the decay of $C_{\n}(r,t)$ in the nematopolar case is faster.
This indicates that nematic structures and orientational singularities are eliminated over longer time scales, resulting in a slower coarsening dynamics.

To quantitatively support this conclusion, in figure~\ref{fig:fig4}(d) we report the typical nematic correlation length $L_{\n}(t)$, estimated as detailed in section~\ref{sec:model_num_meth}, for increasing values of $\bar{k}$.
First, the figure shows that, at sufficiently long times, i.e. when the system enters the ordering regime, all curves follow the scaling law $L_{\n}(t) \sim (t/\ln t)^{1/2}$, in agreement with observations on two-dimensional nematic systems \cite{dutta2005} and XY model \cite{bray2000}.
Moreover, as $\bar{k}$ increases, this scaling trend is preserved, while the absolute magnitude of $L_{\n}(t)$ decreases systematically, supporting our previous conclusion that nematic ordering proceeds more slowly in the nematopolar case.
Similar conclusions are obtained for the length of the polar correlation $L_{\p}(t)$, shown in the inset of figure~\ref{fig:fig4}(d). 
The only relevant difference here is that the curves in the inset tend to overlap, thus showing that $L_{\p}(t)$ is relatively insensitive to the $\bar k$ configuration in which it is obtained.
We emphasize that $L_{\p}(t)$ and $L_{\n}(t)$ represent the characteristic length scales over which the system evolves when probed from the polar or nematic perspectives, respectively.
This is confirmed in the insets of figure~\ref{fig:fig4}(a)–(b), where, rescaling distances by $L_{\p}(t)$ and $L_{\n}(t)$, leads to a collapse of all correlation functions into a single master curve, so that the dynamical scaling hypothesis \cite{bray1994} is satisfied separately in the two perspectives. 
While $L_{\p}(t)$ and $L_{\n}(t)$ are primarily determined by point-like defect dynamics, one should, in principle, also consider an additional scale associated with the evolution of singular extended structures.
This length scale would capture the relaxation of spatial variations in the polarization magnitude $p$. 
However, strings and loops play an important role in the early dynamics, while the late-time evolution is essentially governed by point-like defects.
In the following, we therefore focus on $L_{\p}(t)$ and $L_{\n}(t)$ as the relevant coarsening lengths.

\begin{figure}
 \centering
    \includegraphics[width=0.9\columnwidth]{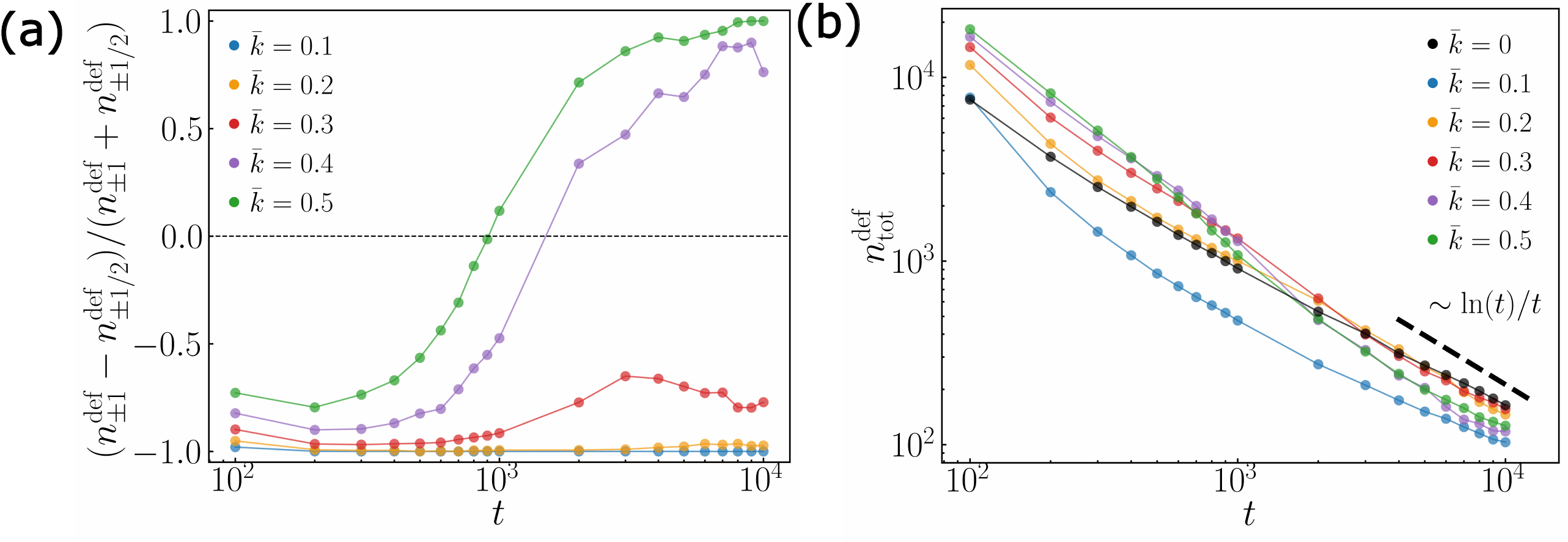}
    \caption{ {\it Defect elimination dynamics in the ordering regime.}
    $(a)$ Time trend of the ratio between the difference and the sum of the number of integer and half-integer defects of both signs in nematopolar configurations for increasing $\bar k$ values.
    $(b)$ Total number of defects $n_{\mathrm{def}}^{\mathrm{tot}}$, given by the sum of integer and half-integer ones, as a function of time for increasing values of $\bar{k}$. 
    As reference, the black line reports the counting in the nematic setting $\bar k=0$.
    The dashed line instead highlights the trend $\sim\ln t/t$.
    All curves are averaged over $5$ independent runs.
    Other parameters are given in section~\ref{sec:num_meth}.
}
\label{fig:fig5}
\end{figure}

Complementary information on the ordering process is obtained by looking at the defect annihilation kinetics in this regime.
In figure~\ref{fig:fig5}(a) we report the time evolution of the ratio between the difference and the sum of the number of defects with integer, $n^{\mathrm{def}}_{\pm1}$, and half-integer, $n^{\mathrm{def}}_{\pm1/2}$, charge, respectively identified and tracked as described in section~\ref{sec:num_meth}.
This curve takes a value $+1$ ($-1$) whenever the system is dominated by integer (half-integer) defects.
Consistently with the predominantly nematic character of the system, in the ordering regime half-integer defects dominate for $\bar{k}=0.1$ and $\bar{k}=0.2$ (ratio $\sim -1$), whereas integer defects become prevalent at $\bar{k}=0.5$ (ratio $\sim +1$), where polar effects are stronger.
The cases $\bar{k}=0.3$ and $\bar{k}=0.4$ report intermediate behaviours, which respectively tend to that of the curves $\bar{k}=0.1$ and $\bar{k}=0.5$ due to the respective still predominant nematic or polar character of the setting.
To extract quantitative information on the global process, in figure~\ref{fig:fig5}(b) we examine the total number of defects $n_{\mathrm{def}}^{\mathrm{tot}}\equiv n^{\mathrm{def}}_{\pm1}+n^{\mathrm{def}}_{\pm1/2}$ for the values of $\bar{k}$ under consideration.
Interestingly, all curves, including the nematic one (black line), are compatible with the scaling behaviour $n_{\mathrm{def}}^{\mathrm{tot}} \sim \ln t/t$, which can be explained by an usual argument as follows.
As mentioned above, if the relevant length scales are $L_{\mathbf p}(t)$ and $L_{\mathbf n}(t)$, which show the same asymptotic scaling, 
it is possible to introduce an effective nematopolar length, $L_{\textbf{np}}(t)\sim(t / \ln t)^{1/2}$, representative of the global degree of order. 
Each defect, on average, occupies an area of order $L_{\textbf{np}}^2(t)$, so that the total number of defects scales as $n_{\mathrm{def}}^{\mathrm{tot}} \sim L^{-2}_{\textbf{np}}(t)$.
Combining this relation with the asymptotic growth of $L_{\mathbf{np}}(t)$ one thus obtains $n_{\mathrm{def}}^{\mathrm{tot}} \sim \ln t/t$, in agreement with the numerical results.

\section{Active nematopolar dynamics}
\label{sec:advection}

Finally, we turn on advection by setting $\Lambda\neq0$ in equation~(\ref{eq:p_equation}), thus making our system active. 
Its presence therefore drives the system out of equilibrium, and its advective character makes it easily identifiable as a self-propulsion contribution which profoundly modifies the overall picture described in section~\ref{sec:ord_dyn_full}.
As a consequence, energetic arguments alone are no longer sufficient to explain the novel behaviours that emerge. 
Rather, it becomes crucial to investigate how the strength and sign of advection affect the stability of the various defect configurations and, consequently, which structures are dynamically selected and persist in the system.

\begin{figure}[t]
    \centering
    \includegraphics[width=0.9\linewidth]{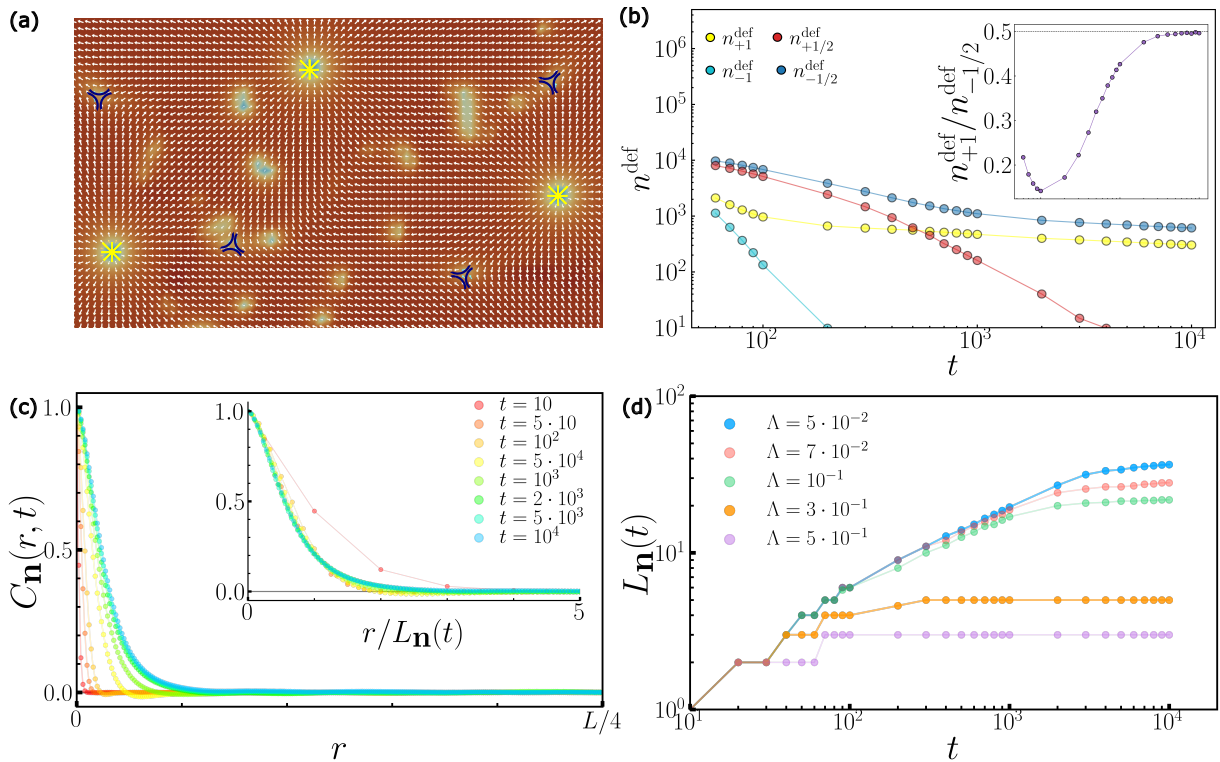}
    \caption{{\it Nematopolar ordering with advection.}
    $(a)$ Magnified view of a representative non-equilibrium nematopolar configuration at $t=5\cdot10^3$ with positive advection strength, started from a disordered initial state and depicted with the same representation style as in figure~\ref{fig:fig1}.
    Here, typical defect structures are trefoils (negative half-integer) and outward-pointing asters (positive integer).
    $(b)$ Time trend of the total number of integer($n^{\mathrm{def}}_{+1}$ and $n^{\mathrm{def}}_{-1}$) and half-integer ($n^{\mathrm{def}}_{+1/2}$ and $n^{\mathrm{def}}_{-1/2}$) defects.
    Inset: time trend of the ratio $n^{\mathrm{def}}_{+1}/n^{\mathrm{def}}_{-1/2}$.
    $(c)$ Nematic correlation function $C_{\n}(r,t)$. 
    Inset: curve collapse obtained by rescaling distances according to the relevant length scale $L_{\n}(t)$ from $(d)$.
    In $(a)$ to $(c)$ data refer to the case $\bar k=0.3$ with advection strength $\Lambda=0.1$.
    $(d)$ Typical nematic length $L_{\n}(t)$ for different values of advection strength $\Lambda$.
    All curves are averaged over $10$ independent runs.
    Other parameters are given in section~\ref{sec:num_meth}.
}
\label{fig:fig6}
\end{figure}

In figure~\ref{fig:fig6}(a) we report a representative configuration for the case $\bar k=0.3$ with $\Lambda=0.1$. 
This shows remarkable qualitative differences with respect to the equilibrium case $\Lambda=0$ shown in figure~\ref{fig:fig1}.
In particular, loops and depolarization strings connecting defects are no longer observed.
Instead, the system is populated almost exclusively by isolated negative half-integer and positive integer defects -- the latter taking the form of outward-pointing asters -- suggesting the idea of a {\it motility-induced charge symmetry breaking (MICS)} mechanism.
Once formed, these structures remain essentially unchanged over long time intervals, suggesting the emergence of a {\it dynamically arrested} regime in which defect evolution is strongly suppressed.

Concerning MICS, its emergence can be qualitatively described in terms of the dynamical stability of defect structures under advection. 
In the case of an outward-pointing aster ($\Lambda>0$), the polarization vectors are oriented opposite to the negative gradient associated with the decrease of the magnitude of polarization towards the defect core.
As a result, the advective term effectively transports polarization from the core, promoting its depletion and thereby stabilizing the structure, while the passive relaxation term tends to restore the equilibrium defect profile, giving rise to a stationary configuration.
By contrast, vortex and spiral configurations are only transient: after an initial rotation driven by the tangential component of the polarization field around the core, they progressively evolve into outward-pointing asters, which represent the dynamically stable integer-defect morphology.
Negative integer defects are instead systematically suppressed. 
Their characteristic quatrefoil structure is incompatible with advection, as polarization is unevenly redistributed among the four arms, destabilizing the defect core. 
Consequently, these defects split and eventually decompose into pairs of stable trefoil structures.
As for positive and negative half-integer defects, these are no longer equivalent under advection, which dynamically selects the trefoil arrangement associated with negative charge. 
Advection, in fact, generates a dynamically consistent circulation of the polarization field around the defect core, thereby allowing these structures to remain stable.
By contrast, comet-like defects disappear as these participate in the mechanism of formation of stable aster defects.

We note that, under the inversion of \(\mathbf p\), which maps an outward aster into an inward one, or under the change of sign of $\Lambda$, the advective term in equation~(\ref{eq:p_equation}) changes sign and is therefore not invariant under these transformations.
In other words, the model exhibits an explicit breaking of symmetry between opposite directions of polarization transport.
As a consequence, when $\Lambda<0$, the dynamically stable integer morphology becomes that of an inward-pointing aster.
Similar flow arguments to those discussed above also apply to half-integer and integer negative defects, with the direction of polarization flux simply reversed.
As a result, trefoil structures remain stable. Comet-like configurations continue to contribute to the formation of positive integer defects, whereas quatrefoil configurations are destabilized and reorganize into trefoil defects.

Support for the idea of an arrested coarsening is provided by figure~\ref{fig:fig6}(b), where we report the time evolution of the number of integer ($n^{\mathrm{def}}_{+1}$ and $n^{\mathrm{def}}_{-1}$) and half-integer ($n^{\mathrm{def}}_{+1/2}$ and $n^{\mathrm{def}}_{-1/2}$) defects, for the advection case with $\Lambda >0$ shown in figure~\ref{fig:fig6}(a). 
After an initial transient, whose duration is comparable to that preceding the onset of the ordering regime in figure~\ref{fig:fig4}(d) for $\Lambda=0$, $n^{\mathrm{def}}_{-1}$ and $n^{\mathrm{def}}_{+1/2}$ decrease very fast, showing that their occurrences are consistently negligible.
In contrast, both $n^{\mathrm{def}}_{+1}$ and $n^{\mathrm{def}}_{-1/2}$ decrease much slower, eventually reaching a plateau regime.
Additionally, the number of negative half-integer defects always exceeds that of positive integer ones.
More quantitatively, the inset reveals that, at late times well inside this plateau regime, $n^{\mathrm{def}}_{+1} \sim 0.5 ~n^{\mathrm{def}}_{-1/2}$, which is fully consistent with the requirement of vanishing total topological charge.

Conclusive reinforcement of this picture is provided by correlation measurements.
In figure~\ref{fig:fig6}(c) we report the nematic correlation function $C_{\mathbf n}(r,t)$, sampled at the same times as in figure~\ref{fig:fig4}(a)–(c), once again for the case explored in figure~\ref{fig:fig6}(a).
The figure clearly shows that, once the system enters the late-time plateau regime illustrated in figure~\ref{fig:fig6}(b), the correlation functions at successive times exhibit very similar decay and nearly overlap.
This behaviour is further clarified in figure~\ref{fig:fig6}(d), where we report the typical nematic length scale $L_{\mathbf n}(t)$ for different values of the positive advection strength at fixed $\bar k=0.3$. 
Interestingly, also in this case all curves eventually reach a plateau, indicating that the coarsening is indeed arrested. 
In particular, increasing $\Lambda$ leads to an earlier arrest of coarsening, resulting in a larger number of stationary defects at late times and, correspondingly, smaller inter-defect separations, as reflected by the reduced plateau values of the characteristic length scale.
For completeness, as shown in the inset of figure~\ref{fig:fig6}(d), we mention that the dynamical scaling hypothesis remains valid when distances are rescaled using these plateauing $L_{\n}(t)$.

\section{Conclusions and Discussion}
\label{sec:conclusions}

Active systems displaying both polar and nematic alignment are often found in nature \cite{volfson2008, zhou2014, doostmohammadi2016, genkin2017, sokolov2015, kawaguchi2017, saw2017, blanch2018, Turiv2020, meacock2021, ruider2024, wheeler2024, han2025, ma2026} and provide a fertile theoretical framework for investigating how competing symmetries give rise to defect structures and ordering mechanisms without counterpart in purely polar or nematic systems \cite{amiri2022, vats2024, vafa2025, mishra2025, dinelli2026}.

In this work, we numerically investigated phase ordering and defect dynamics in a newly introduced minimal coarse-grained model for dry nematopolar systems. 
Unlike previous nematopolar models, our approach involves a single polar field, with polar and nematic interactions arising from competing contributions to the free energy. 
Activity is instead incorporated through a non-equilibrium self-advection term.

We showed that, when polar and nematic alignments are optimally balanced, the system develops peculiar structures absent in purely polar or nematic settings, namely depolarization lines connecting half-integer defects and separating domains of opposite polarization, and closed depolarization loops, which together act as preferred pathways for energy minimization.
Through initial controlled simulations, we characterized the elementary relaxation mechanisms of these structures and elucidated the interplay between polar and nematic alignment. 
Oppositely charged defects move along the connecting string until annihilation, with an attractive interaction depending on the relative strengths of the polar and nematic couplings.
Like-charged defects reach a finite monotonically-varying equilibrium separation resulting from the balance between repulsive and string-mediated interactions. 
Finally, depolarization loops are found to disappear through two distinct mechanisms: evaporation, favoured when nematic order dominates and loops are thicker, and polarization reorientation, in the opposite regime.
Overall, these findings confirm and extend findings from previous nematopolar settings \cite{amiri2022, vafa2025, mishra2025, dinelli2026, ma2026}.

Starting from disordered initial conditions, we found that the system orders verifying dynamical scaling, with a nematic length scale growing as $\sim (t/\ln t)^{1/2}$ consistent with ordering dynamics governed by point-like defects.
This scaling law is confirmed by complementary measurements of the decay trend of the number of defects observed during system evolution.

We further showed that self-advection strongly alters this scenario. 
At sufficiently high advection, the activity induces motility-induced charge symmetry breaking between defect species, preferentially stabilizing positive integer and negative half-integer defects. 
As a consequence, a finite defect density persists at long times, the characteristic length scale saturates, and the system enters an arrested-coarsening regime.

Overall, our work demonstrates that a simple and minimal single-field description is already sufficient to reproduce a wealth of unconventional defect structures and ordering phenomena in nematopolar systems.
In particular, it highlights activity as a fundamental mechanism capable of reshaping defect-mediated ordering and stabilizing non-equilibrium states that have no counterpart in passive systems. 
Therefore, we expect that this framework will provide a useful starting point for investigating a broader class of active materials with competing symmetries, while offering a unified perspective for interpreting the complex behaviour observed in biological and synthetic nematopolar systems such as ferroelectric liquid crystals \cite{chen2020, lavrentovich2020, basnet2022, kumari2023, ma2024}, living liquid crystals \cite{zhou2014, genkin2017, sokolov2015, Turiv2020}, bacteria \cite{volfson2008, doostmohammadi2016, meacock2021, wheeler2024, han2025} and eukaryotic cells \cite{saw2017, kawaguchi2017, blanch2018, ruider2024, ma2026} colonies and microtubule–motor mixtures \cite{kruse2005, sumino2012, huber2018, roostalu2018}. 
A further insightful biological example is that of the {\it Hydra} polyp \cite{wang2023}, where topological positive integer and negative half-integer defects in muscle fiber orientation have been shown to localize to key features of the body plan.

Natural extensions of the present work inspired by recent experiments \cite{comba2022, eckert2023, li2025} may include generalizations in which the free energy incorporates competing symmetries beyond the polar and nematic ones, together with symmetry-constrained advective contributions and additional physical ingredients such as density fluctuations or thermal noise. 
Exploring these directions may reveal whether the mechanisms identified here -- especially in the active nematopolar setting -- represent generic features of systems with competing orientational orders.

\section*{Acknowledgements}
Numerical calculations have been made possible through a Cineca-INFN agreement, providing access to HPC resources at Cineca. All authors acknowledge support from INFN/FIELDTURB project and from MUR projects Quantum Sensing and Modelling for One-Health (QuaSiModO). This research was supported in part by grant NSF PHY-2309135 to the Kavli Institute for Theoretical Physics (KITP). M.S. thanks A. Maitra for insightful discussions.

\section*{Data availability statement}
All data generated for the investigation are reported in the paper and can be made available upon reasonable request to the corresponding author.

\section*{Author contributions}
All authors contributed equally to conceptualization, numerical investigation, writing, figure generation and general supervising of this paper.

\appendix
\section{Adimensional equations and natural time unit}
\label{app:adim_eqs}

It is common practice to rescale the equation of motion equation~(\ref{eq:p_equation}) in dimensionless units.
Similarly to \cite{cahn1958, gunton1990, bray1994, wittkowski2014}, such an equation can be recast by a suitable rescaling of length, time and fields. Simple steps reveal that the relevant time unit is $\tau_{\p}=1/(\Gamma \alpha_{\p})$.
Since this is the unique time unit which emerges from manipulation of equation~(\ref{eq:p_equation}), we identify it as a natural time unit $\tnat\equiv\tau_{\p}$.

Concerning the length unit, we instead have two possible choices, either $\ell_{\p}=\sqrt{k_{\p}/\alpha_{\p}}$ or $\ell_{\n}=\sqrt{k_{\n}/\alpha_{\p}}$, which emerge as the natural length units for purely polar and nematic systems, respectively. 
We anticipate that, in order to avoid confusion on the choice of natural length unit in our nematopolar setting, we extract it from the decay of the polarization magnitude sufficiently far from the core of a defect (see appendix~\ref{app:pol_dec}). 
For the two length units at hand, we respectively get
\begin{equation}
\begin{split}
    \frac{d\bar\p}{d\bar{t}}+\bar\Lambda_{\p}((\bar\p\cdot\nabla)\p) &= -\left( |\bar\p|^2 \bar\p - \bar\p - \bar{\nabla}^2 \bar\p\right. \\
                                                                     &\left.-\frac{2}{\bar k}\left[
                                                                      (\bar\p\cdot\bar\nabla^2 \bar\p)\bar\p
                                                                      + 2\bar\p\cdot\big((\bar\nabla\otimes\bar\p)^T(\bar\nabla\otimes\bar\p)\big)
                                                                      + \bar p^2\,\bar\nabla^2 \bar\p\right] \right)
\end{split}
\label{eq_p_resc}
\end{equation}
or
\begin{equation}
\begin{split}
    \frac{d\bar\p}{d\bar{t}} +\bar\Lambda_{n}((\bar\p\cdot\nabla)\p &= -\left( |\bar\p|^2 \bar\p - \bar\p - \bar k \bar{\nabla}^2 \bar\p\right. \\
                                                                    &-2\left.\left[
                                                                     (\bar\p\cdot\bar\nabla^2 \bar\p)\bar\p
                                                                     + 2\bar\p\cdot\big((\bar\nabla\otimes\bar\p)^T(\bar\nabla\otimes\bar\p)\big)
                                                                     + \bar p^2\bar\nabla^2 \bar\p\right]\right)
\end{split}~,
\end{equation}
where the bar symbol denotes dimensionless quantities, 
\begin{equation}
    \bar k \equiv \frac{k_{\p}}{k_{\n}}~,\qquad \bar\Lambda_{\p}\equiv\frac{\Lambda}{\Gamma\sqrt{\alpha_{\p}k_{\p}}}~,\qquad \bar\Lambda_{\n}\equiv\frac{\Lambda}{\Gamma\sqrt{\alpha_{\p}k_{\n}}}
\end{equation}
are the only surviving adimensional parameters and the polarization field is rescaled as $\bar\p=\p/\p_\text{resc}$, with $\p_\text{resc}=1$ unitary value of polarization at equilibrium. 
From equation~(\ref{eq_p_resc}), note that, whenever the nematic and self-advective contributions are absent ($\Lambda_{\p}=k_{\n}=0$), the polar-like polarization equation with all parameters equal to one is recovered.

\section{Polarization field decay and natural length unit}
\label{app:pol_dec}

In this section we provide an approximate expression for the magnitude of $\p$ sufficiently far from the core of a defect of generic charge $q$. 
To do this, we recast our problem, i.e. free energy equation~(\ref{eq:free_energy}) and equation of motion equation~(\ref{eq:p_equation}), in polar coordinates $(r,\phi)$, $r\in(0,+\infty]$, $\phi\in[0,2\pi)$, and exploit the radial structure of $\p$ around defect cores.

To start, we recast the polar vector as
\begin{equation}
    \p \equiv p \mathbf{n}=p(\cos\theta, \sin \theta)~,
\end{equation}
 where $p\equiv p(r)$ and $\theta\equiv\theta(r)$ respectively are polarization magnitude and orientation, and $\mathbf{n}$ is the unit orientation field vector. 
 From this, we evaluate $(\nabla \mathbf{p})^2$ and $(\nabla \hat{P})^2$.
 We remark that $\mathbf{n}$ rotates $q$ times faster than the polar frame, therefore $\mathbf{n}$ coincides with the radial unit vector only for $q=1$. 
 Concerning $(\nabla \mathbf{p})^2$, using that $\partial_i \enne=\enne^\perp \partial_i \theta$, with $\enne^\perp$ unit vector orthogonal to $\enne$, we readily get
 \begin{equation}
    (\nabla \mathbf{p})^2=(\nabla p)^2+p^2(\nabla \theta)^2~.
\label{eq:grad_p_vec}
\end{equation}
Concerning $(\nabla \hat{P})^2$, we first recast the nematic tensor equation~(\ref{eq:nematic_tens}) as $\hat{P}=p^2 \hat A$ with $A_{ij}=(n_i n_j -\delta_{ij}/2)$.
Then, using 
$\partial_i \hat A=(\enne^\perp \enne + \enne \enne^\perp)\partial_i \theta$ and also that
\begin{equation}
    A_{jk}A_{jk}=\frac{1}{2}, \quad
    A_{jk}\partial_iA_{jk}=0, \quad
    \partial_iA_{jk}\partial_iA_{jk}=2(\partial_i\theta)^2~,
\end{equation}
we obtain
\begin{equation}
    (\nabla \hat{P})^2=
    2p^2(\nabla p)^2+2p^4 (\nabla \theta)^2~.
\label{eq:grad_p_tens}
\end{equation}

Having determined the expressions for elastic penalties, we are now ready to express the free energy for an isolated defect of general charge $q$. 
For such a defect, the angular profile is given by $\theta=q\phi$, whence one easily gets that $(\nabla \theta)^2=q^2/r^2$ and $ (\nabla p)^2=(p')^2$ ($p'\equiv dp/dr$). 
As a consequence, the free energy takes the form
\begin{equation}
    F[p]=
    \int rdr d\phi \left[
    -\ap\left( -\frac{p^4}{4} + \frac{p^2}{2}\right)
    + \frac{1}{2}(\kp +2\kn p^2)(p')^2
    + \frac{1}{2}\frac{q^2}{r^2}(\kp +2\kn p^2)p^2
    \right]~,
\label{eq:free_en_polar_coords}
\end{equation}
whence we get that the equation of motion is 
\begin{equation}
    \ap(p^3-p)
    -2\kn p(p')^2
    +\frac{q^2}{r^2}(\kp+4\kn p^2)p
    -(\kp+2\kn p^2)
    \left(
    \frac{p'}{r}+p''
    \right)
    =0~.
\label{eq:p_equation_radial}
\end{equation}

Sufficiently far from the origin the polarization magnitude can be expressed as
\begin{equation}
    p=1-\delta p,\qquad\delta p\ll 1~.
\end{equation}
Introducing this expression in equation~(\ref{eq:p_equation_radial}) and linearizing in $\delta p$, we finally obtain
\begin{equation}
    \delta p'' =
    \frac{2\ap}{\kp+2\kn}\delta p~.
\label{eq:deltap_equation}
\end{equation}
where we used that that for $r\gg1$, where $\delta p \ll 1$, contributions $\sim 1/r^n,~n\in\mathbb{N}$ can be neglected.
Recalling that $p\rightarrow1$ as $r\uparrow\infty$, equation~(\ref{eq:deltap_equation}) is solved by 
\begin{equation}
    \delta p\sim e^{-\frac{\sqrt{2}}{\lnat}r}, \qquad\text{with}\qquad \lnat\equiv\sqrt{\frac{\kp+2\kn}{\ap}}~,
\end{equation}
where $\lnat$ is a length that regulates the decay of $p$ sufficiently far from the origin, and that we therefore identify as our natural length unit.

\section*{Bibliography}

\bibliography{Bibliography_arxiv.bib}

\end{document}